\documentclass[letterpaper,12pt]{extarticle}
\usepackage{tabularx} 
\usepackage{preamble}
\usepackage{packages}
\usepackage{breqn}
\usepackage{enumerate}
\usepackage[font=small,labelfont=bf]{caption}
\usepackage[left=20mm,right=20mm,top=30mm,bottom=30mm]{geometry}
\usepackage{pgfplots}
\usetikzlibrary{calc}
\usepackage{extarrows}
\usepackage{amsmath,amssymb}
\usepackage{amsfonts}
\usepackage{bbm}
\usepackage{booktabs,subcaption,amsfonts,dcolumn}
\newcolumntype{d}[1]{D..{#1}}
\definecolor{refkey}{rgb}{0.9451,0.2706,0.4941}
\definecolor{labelkey}{rgb}{0.9451,0.2706,0.4941}
\DeclareFontFamily{U}{rcjhbltx}{}
\DeclareFontShape{U}{rcjhbltx}{m}{n}{<->rcjhbltx}{}
\DeclareSymbolFont{hebrewletters}{U}{rcjhbltx}{m}{n}
\DeclareMathSymbol{\tet}{\mathord}{hebrewletters}{84}
\DeclareMathSymbol{\pey}{\mathord}{hebrewletters}{112}
\def\muchlessthan{\ll}
\def\muchgreaterthan{\gg}
\newcommand{\kakko}[1]{\left(#1\right)}
\def\z2{$\mathbb{Z}_2$}

\definecolor{darkgray}{rgb}{0.33, 0.33, 0.33}
\usepackage{braket}
\newcommand{\kket}[1]{\left|\left.#1\right\rangle\!\right\rangle}

\newcommand{\bbra}[1]{\left\langle\!\left\langle#1\right|\right.}
\newcommand{\Tr}{{\mathrm{Tr}}}

\DeclareMathOperator{\sgn}{sgn}
\usepackage{authblk}
\def\phibar{\bar\phi}
\definecolor{darkgray}{rgb}{0.33, 0.33, 0.33}

\title{\Large Chern-Simons-matter theories at large baryon number}
\author[1,2,3]{Masataka Watanabe}
\affil[1]{\small Albert Einstein Center for Fundamental Physics, Institute for Theoretical Physics, University of Bern,
Sidlerstrasse 5, CH-3012 Bern, Switzerland}
\affil[2]{\small Kavli Institute for the Physics and Mathematics of the Universe (WPI), The University of Tokyo Institutes for Advanced Study, The University of Tokyo, Kashiwa, Chiba 277-8583, Japan}
\affil[3]{\small Department of Physics, Faculty of Science, The University of Tokyo, Bunkyo-ku, Tokyo 133-0022, Japan}
\date{}
\preprint{{\tt IPMU-19-0061}}

\allowdisplaybreaks[4]

\begin{document}

  \maketitle 
  
  \begin{abstract}
      We study $SU(2)$ Chern-Simons theories at level $k$ coupled to a scalar on $T^2\times \mathbb{R}$ at large baryon number.
We find a homogeneous but anisotropic ground state configuration for any values of $k$ on the IR fixed-point of those models.
This classical analysis is valid as long as we take the baryon number large.
As a corollary, by comparing the symmetry breaking pattern at large chemical potential, we find that the theory does not reduce to the singlet sector of the $O(4)$ Wilson-Fisher fixed-point at large-$k$, as expected from general grounds.
This paper will be one primitive step towards quantitative analysis of Chern-Simons-matter dualities using the large charge expansion.



  \end{abstract}
      \newpage
\setcounter{tocdepth}{2}
\setcounter{secnumdepth}{4}
\tableofcontents
\hypersetup{linkcolor=PaleGreen4}
\newpage

\section{Introduction}
\label{sec:intro}


It is a generic feature of quantum field theory with a symmetry to simplify at large global charge.
In the sector of large global charge, asymptotic expansions of the effective Lagrangian are often possible, enabling us to compute the operator dimensions and OPE coefficients to any given order, using a small number of undetermined coefficients in front of terms allowed by symmetries of the system~\cite{Hellerman:2015nra,Alvarez-Gaume:2016vff,Monin:2016jmo}.
In~\cite{Hellerman:2015nra,Alvarez-Gaume:2016vff,Monin:2016jmo}, the lowest dimensions of operators at large charge were computed by writing down such effective Lagrangians on a spherical spatial slice $S^{D-1}$, \it via \rm the state-operator correspondence.
The result for the lowest operator dimension was later verified numerically in \cite{Banerjee:2017fcx,delaFuente:2018qwv,Banerjee:2019jpw}.\footnote{For $AdS$ interpretation of such a result see \cite{Loukas:2018zjh}.}

The reason why such an effective action is valid is because the effective Lagrangian has its UV scale at $\Lambda_{\rm UV}\equiv {\rho}^{\frac{1}{D-1}}$ and IR at $\Lambda_{\rm IR}\equiv 1/R_{\rm geometry}$ if we consider a system with large global charge $J$ and the charge density $\rho$:
The large hierarchy of scales, $\Lambda_{\rm IR}/\Lambda_{\rm UV}\propto J^{-\frac{1}{D-1}}$, suppresses both quantum effects and higher-derivative terms and renders the theory semi-classical and weakly-coupled, similar to the method of chiral Lagrangian.\footnote{When the original theory has a moduli space of vacua, the situation is a bit different, in which $\Lambda_{\rm UV}\propto \sqrt{J}/R_{\rm geometry}$. This is because the spectrum of such a theory is infinitely degenerate in the flat space limit~\cite{Hellerman:2017veg,Hellerman:2017sur,Hellerman:2018xpi}.}
The argument is true in any reasonable finite-sized spatial slices, for example on the torus where we can compute the ground state energy in the presence of the chemical potential.

However, one important assumption, which can sometimes be proven on a case-by-case basis, is actually made in the last paragraph --
The ground state of such a Lagrangian must be homogeneous, or inhomogeneous but at the IR scale.
This is because if the ground state configuration computed from the effective Lagrangian is inhomogeneous at the UV scale, $\Lambda_{\rm UV}\equiv {\rho}^{\frac{1}{D-1}}$, one should not have used the effective Lagrangian in the first place.

Inhomogeneous ground states were first shown to exist in \cite{Alvarez-Gaume:2016vff}, where examples were explicitly constructed in \cite{Hellerman:2017efx,Hellerman:2018sjf} for the $O(4)$ Wilson-Fisher fixed-point, proving the inhomogeneity to be at the IR scale for this model.
No examples of relativistic CFTs with a UV inhomogeneous ground state at large charge is known up until now (although free fermion and its dual Chern-Simons-matter theory is a strong candidate for this).
For a list of rough classifications (IR-inhomogeneity, UV-inhomogeneity, etc.) of phases of matter realised in the large charge limit, see~\cite{Hellerman:2018sjf}.
Consult \cite{Hellerman:2015nra, Alvarez-Gaume:2016vff, Loukas:2017lof, Loukas:2017hic, Hellerman:2017efx,Hellerman:2018sjf} and \cite{Favrod:2018xov, Kravec:2018qnu, Kravec:2019djc} for known phases actually realised in various relativistic and non-relativistic systems at large charges, respectively.

Not only is it important in checking the consistency of the effective field theory approach, studying the ground state and its symmetry breaking pattern at large global charge is also useful in classifying the various phase of matter realised at large charge/chemical potential.
The candidates of phases realised at large global charge known so far, probably among others, are
\begin{enumerate}[(1)]
    \item {\bf Axio-dilaton phase}: The effective theory at leading order becomes free, such as SUSY theories with moduli space of vacua \cite{Hellerman:2017veg,Hellerman:2017sur,Hellerman:2018xpi,Hellerman:2020sqj}, or simple the free scalar theory.\footnote{The former is slightly different from the latter in that the fermion is also massless, but we neglect this here.}
    \item {\bf Fermi sea}: The theory forms a Fermi sea at large charge, as realised in free fermion.
    \item {\bf Superfluid phase}: The effective action at large charge contains one NG mode, $\chi\sim \chi+2\pi$, as realised in $O(2)$ Wilson-Fisher fixed-point in three dimensions \cite{Hellerman:2015nra}, or $D=3$, $\mathcal{N}=2$ theory with superpotential $W=\Phi^3$ \cite{Hellerman:2015nra}.
    \item{\bf FFLO phase (Inhomogeneous superfluid phase)}: Typically realised in examples with non-Abelian symmetry and with two chemical potentials turned on, the ground state breaks the translation invariance of the spatial slice. Examples include $O(4)$ Wilson-Fisher theory with two chemical potentials excited.
    The name ``Fulde–Ferrell–Larkin–Ovchinnikov (FFLO) phase'' comes from the phase found in \cite{Larkin:1964wok, Fulde:1964zz} where the chiral condensate of superconductors varies in space.
\end{enumerate}
We call these phases of matter the ``the large charge universality class,'' which is a more rough classification of the ordinary universality classes.

The purpose of this paper is to study further these interesting phases of matter at large baryon number density, $\rho_B$.
In this paper, we will find a new phase of matter at large global charge, which is \it a homogeneous superfluid phase with rotational symmetry broken on the spatial slice. \rm 
The model we consider here is the three-dimensional $SU(2)$ Chern-Simons theory at level $k$, coupled to a single boson with a quartic or sextic potential.\footnote{
It is believed that there is an infrared fixed point for this model \cite{Hsin:2016blu}, where one can find a dual fermion model with correct anomalies. 
The beta function of $SO(N)$ Chern-Simons-matter theory~\cite{Aharony:2011jz} can be also explicitly computed, and it is known that there is a weakly-coupled fixed-point for $N\geq 10$.
If, for whatever reason, it turns out that the assumption about the infrared fixed point is false, we can just replace $h^2$ with $h^2[\Lambda]$, and think about the one-loop running of it.
}
We will first write down the effective action of this at large charge, and then proceed to study its ground state configuration.
For simplicity, we will only think about the torus spatial slice and not the spherical one, and there 
we will argue that such a configuration breaks the rotational symmetry spontaneously.

Aside from finding a new phase of matter at large global charge,
this paper is the first one to show that a fully-controlled analysis of strongly-coupled Chern-Simons-matter theories is possible at finite $N$, when we use the effective field theory approach.\footnote{Nonetheless, a paper similar in spirit to ours exists~\cite{Kumar:2018nkf}, which studied the $SU(N)$ Chern-Simons-matter theory at finite density at the level of the classical Lagrangian, in the weak coupling regime (in contrast to what we did, which is a fully controlled calculation even at strong coupling).
}
Also, the regime of finite-$N$ and large baryon number is a completely new regime of study in the study of Chern-Simons-matter theories and their dualities.
Combined with \cite{1904.07885}, which studies various large-$N$ Chern-Simons-matter theories at finite chemical potential, we can hope to unravel more interesting aspects of their dualities.


Related, one practical aspect of the large charge universality class is that we can check the dualities by checking if two theories belong to the same class.
The final goal in this regard should be checking the bose-fermi duality relating Chern-Simons theories coupled to bosons and fermions, by writing down the effective theory on both sides and comparing the two.
The analysis of systems including fermions have seldom been considered at large charge,\footnote{
Models with supersymmetry (hence containing fermions) with or without moduli have been considered in \cite{Hellerman:2017sur,Hellerman:2018sjf,Hellerman:2018xpi} and in \cite{Hellerman:2015nra}, respectively.
There have been studies about non-relativistic models containing fermions too \cite{Favrod:2018xov,Kravec:2018qnu, Kravec:2019djc}.
}
but this could become one step forward towards such an analysis, with an ultimate goal of understanding particle-vortex dualities \cite{Karch:2016sxi, Seiberg:2016gmd, Murugan:2016zal} in a fully controlled, quantitative fashion.

As a baby version of the problem, we start by checking if the theory in question reduces to the singlet sector of $O(4)$ Wilson-Fisher fixed-point at $k\to \infty$ in this paper, which is a common folklore in studying Chern-Simons-matter theories.
To make things complicated, this statement is indeed true on the spherical spatial slice but not on other spatial slices, in particular the torus, if we think about it carefully \cite{Banerjee:2012gh, Banerjee:2013mca}. 
We will review this in the main body of the text, and then see the fact that this statement is false on the torus in a different way than \cite{Banerjee:2012gh, Banerjee:2013mca} --  we compare the ground state configuration and its symmetry breaking pattern of $SU(2)_k$ Chern-Simons theory coupled to fundamental matter, with that of the $O(4)$ Wilson-Fisher theory projected onto the singlet sector.
As the former does not break the translation invariance while the latter does, we will conclude that the common folklore is false on the torus spatial slice, as expected from \cite{Banerjee:2012gh, Banerjee:2013mca}.

The rest of the paper is organised as follows.
In Section \ref{sec:ground}, we will compute the ground state configuration of the Chern-Simons-matter theory at large baryon number, and show that it is homogeneous but anisotropic, which is the main result of the paper.
In Section \ref{consequence}, we discuss the implications of the result.
We first discuss the regime of validity of the ground state configuration, and see that it is valid for all values of $k$, incluidng $k\to \infty$.
We then use this to rederive that the theory does not reduce to the singlet sector of $O(4)$ Wilson-Fisher fixed-point, on the torus spatial slice.
Finally in Section \ref{conc}, we will conclude the paper with a summary and possible interesting future directions.


\section{Chern-Simons-matter theory at large baryon number}
\label{sec:ground}

\subsection{$SU(2)_k$ Chern-Simons theory coupled to fundamental scalar}

\subsubsection{Two models and the effective action}

There are two conformal field theories we are interested in in this paper.
One is the $O(4)\sim SU(2)\times SU(2)$ Wilson-Fisher scalar $\Phi$ whose diagonal $SU(2)$ is gauged and coupled to an $SU(2)$ Chern-Simons theory at level $k$.
In the language of \cite{1808.03317,1808.04415,1904.07885,2008.00024}, this is the $SU(2)_k$ critical boson theory, which is conjectured to be dual to $U(k)_{-3/2}$ regular fermion theory (See also \cite{Hsin:2016blu}).
The theory is defined by the UV Lagrangian
\begin{eqnarray}
S_{\mathrm{CSM}} &\equiv& S_{\rm CS}+S_{\rm boson}\\
S_{\rm CS} & \equiv& \frac{k}{4\pi}
\int
 \Tr \kakko{
A \wedge dA - \frac{2i}{3} A\wedge A\wedge A
}
\\
&=&\frac{k}{4\pi}\int d^3x\, \epsilon^{\mu\nu\rho}\Tr\kakko{A_\mu\partial_\nu A_\rho +\frac{2i}{3}A_\mu A_\nu A_\rho}
\\
S_{\rm critical} &\equiv &\int d^3x
\kakko{\partial_\mu +iA_{\mu}}\bar\Phi \kakko{\partial^\mu -iA^{\mu}} \Phi + g_4  \kakko{\bar\Phi\Phi}^2.
\end{eqnarray}
We used the Lorentzian signature with mostly plus metric, $g_{\mu\nu}=\kakko{-,+,+}$.
For later reference, we also define the gauge field strength as
\begin{equation}
F_{\mu\nu}\equiv \partial_\mu A_\nu-\partial_\nu A_\mu 
+i[{A_\mu},{A_\nu}].
\end{equation}
It is believed that such a model has an infrared fixed-point (Indeed, the large-$N$ beta functions are computed in \cite{1808.03317} to show this.), and it is this fixed-point that we are going to study in this paper.

The other theory, which is also to believed to flow to an infrared fixed-point, is defined by the UV Lagrangian by replacing $S_{\rm critical}$ by
\begin{align}
    S_{\rm regular} &\equiv \int d^3x
\kakko{\partial_\mu +iA_{\mu}}\bar\Phi \kakko{\partial^\mu -iA^{\mu}} \Phi + g_6  \kakko{\bar\Phi\Phi}^3,
\end{align}
which is called the regular boson theory, which is likewise conjectured to be dual to the critical fermion theory \cite{1808.03317}.
This theory is obtained by tuning the relevant coupling $g_4$ to zero, and the large-$N$ computation (of $SO(N)$ Chern-Simons matter theory) shows that $g_4$ scales as $1/k$ at large $k$ \cite{Aharony:2011jz}.
This is most likely true for our case as well.

Now we put the theory on the torus and turn on the chemical potential for the $U(1)$ baryon symmetry.
One can then write down the effective field theory around the saddle point of the Lagrangian with the chemical potential added, which is, as was demonstrated in \cite{Hellerman:2015nra}, composed of effective operators of dimension $3$ which is invariant under all the symmetries of the system.
The Lagrangian at leading order at large chemical potential therefore becomes
\begin{eqnarray}
S_{\mathrm{eff}} &\equiv& S_{\rm CS}+\int d^3x \left[
\kakko{\partial_\mu +iA_{\mu}}\phibar \kakko{\partial^\mu -iA^{\mu}} \phi + \frac{h^2}{12}  \kakko{\phibar\phi}^3\right],
\label{eq:EFT}
\end{eqnarray}
where $h$ is in general $O(1)$.
We will therefore not see the difference between critical boson theory and the regular boson theory at leading order in the large charge expansion, except for the possible difference for the values of $h$.

A few comments are in order about this effective action.
First, the field $\phi$ is different from $\Phi$ which was used in the UV Lagrangian, and must be thought of as an axio-dilaton, as explained in \cite{2008.03308}.
Second, the Chern-Simons term does not get renormalised, since the level is quantised and we do not integrate fermions out to obtain the effective action.
Lastly, since we consider the torus spatial slice, we do not need to add any conformal coupling to make the action Weyl-invariant.

\subsubsection{The leading order equations of motion}

We now would like to write down the leading order equations of motion (EOMs) for this system, which will be used later on.
The equation of motion for $\phi$ is simply
\begin{equation}
D_\mu D^\mu \phi-\frac{h^2}{4}\left(\bar{\phi}\phi\right)^2 \phi
=0
\label{eom}
\end{equation}
We also have the Gauss law constraint, which is nothing but the equation of motion for the gauge fields,
\begin{equation}
-\frac{k}{4\pi}\kakko{\ast F^{(a)}}^\mu=
-i\bar\phi_i{\sf T}^a_{ij}D^\mu\phi_j+i\kakko{D^\mu\bar\phi_i}{\sf T}^a_{ij}\phi_j,
\label{Gauss}
\end{equation}
where $\ast$ denotes the Hodge star operator, acting on $F_{\mu\nu}$ as
\begin{equation}
\kakko{\ast F^{(a)}}^\mu = \frac12\sqrt{\abs g} g^{\mu\nu}\varepsilon_{\nu\rho\sigma}F^{\rho\sigma, (a)},
\end{equation}
while $T^{(a)}\equiv \sigma^{(a)}/2$ denotes the generators of $SU(2)$ in the fundamental representation.

\subsubsection{Taking the large baryon number density}

By virtue of the Noether theorem, the baryon number density of this model is nothing but the RHS of \eqref{Gauss} with $T$ replaced by $\mathbbm{1}$,
\begin{equation}
J^{\mu}_B=-i\phibar \kakko{D^\mu \phi}+i\kakko{D^\mu\phibar}\phi.
\end{equation}
Therefore, the charge density of this model becomes
\begin{equation}
\rho_B=-i\phibar \kakko{D^0 \phi}+i\kakko{D^0\phibar}\phi.
\label{chargedensity}
\end{equation}
The spatial integral $J_B$ of this quantity is what we are going to take large in this paper.

There also are two global currents $J^\mu_1$ and $J^\mu_2$, which are the generators of the global $SO(3)$ symmetry along with $J_B^\mu$,
\begin{eqnarray}
J^\mu_+=2i\epsilon_{ab}\phi^a\kakko{D^\mu \phi}^b, \qquad
J^\mu_-=-2i\epsilon_{ab}\phibar^a\kakko{D^\mu \phibar}^b,
\label{twoother}
\end{eqnarray}
where we have organized as $J^\mu_{\pm}\equiv J^\mu_1\pm iJ^\mu_2$.
We fix the charge associated to these currents to vanishing, however.

\subsection{Ground state solution at large charge on the torus}

\subsubsection{Computing the ground state configuration at large baryon number}

We now look for the ground state solution to the above set of EOMs, on the torus.
In this regime, it is most convenient to take the (helical) unitary gauge,
\begin{equation}
\phi\equiv
\begin{pmatrix}
\abs{q}e^{i\omega t}\\
0
\end{pmatrix}
\equiv \begin{pmatrix}
ve^{i\omega t}\\
0
\end{pmatrix}
\end{equation}
We now make an ansatz that the ground state configuration is homogeneous, and if we find such a solution, it will have the lowest energy. 
This assumption simplifies the EOMs a lot.\footnote{
I used Mathematica to solve the simplified EOMs, and the file can be prepared upon request. Meanwhile one can check it by hand because the computation is not really that complicated.
}

We first fix the baryon number density,
\begin{equation}
\rho_B=-v^2\kakko{A^{0,(3)}-2\omega}.
\end{equation}
According to the homogeneous ansatz, the Gauss law constraint \eqref{Gauss} while fixing the baryon number density can be solved as, up to spatial rotations to get \eqref{eq:A} and \eqref{eq:omega}, and afterwards we can use the EOM for the matter field \eqref{eom} to solve for $v$, as in \eqref{eq:v4}.
The final result for the ground state configuration looks like
\begin{align}
\begin{split}
\begin{pmatrix}
A^t, & A^x, & A^y
\end{pmatrix}
&=
\begin{pmatrix}
-\frac{2\pi v^2}{\abs{k}} \times T^{(3)},&
\sqrt{\frac{2{\pi\rho_B}}{\abs{k}}}\times T^{(1)},&
\sgn(k)\sqrt{\frac{2{\pi\rho_B}}{\abs{k}}}\times T^{(2)}
\end{pmatrix}\\
\phi&\equiv
\begin{pmatrix}
\abs{q}e^{i\omega t}\\
0
\end{pmatrix}
\equiv \begin{pmatrix}
ve^{i\omega t}\\
0
\end{pmatrix}
\end{split}
\label{eq:vev}
\end{align}
while
\begin{eqnarray}
\omega&=&-\frac{\pi v^2}{\abs{k}}+\frac{\rho_B}{2v^2}
\label{eq:omega}\\
v^4&=&\frac{\rho_B}{h^2\abs{k}}\kakko{\sqrt{h^2k^2+4\pi^2}-2\pi}>0
\label{eq:v4}
\end{eqnarray}
Note that the above expressions respects the symmetry of inverting $k$ and $x_2$ at the same time.

\subsection{Spontaneous symmetry breaking at large baryon number}


Let us discuss the property of the configuration we computed above.
First of all, the configuration is homogeneous but anisotropic, i.e., it breaks the rotational symmetry on the spatial slice.
Even though this fact is almost clear by looking at the ground state configuration, in order to characterise the anisotropy in a gauge invariant way, we will use the other current than the baryon number current, of the global $SO(3)$ symmetry, defined in \eqref{twoother}.

By plugging in the ground state configuration to \eqref{twoother}, we get\footnote{Although this is indeed an example of persistent current and seems to violate the Block's theorem,
\cite{Tada2016, doi:10.1143/JPSJ.65.3254, 2019arXiv190402700W, PhysRev.75.502, PhysRevD.92.085011, PhysRevB.78.144404}, in fact it does not.
The Bloch theorem states that there is no persistent current for a ground state, but in order to make our configuration a precise ground state, we need to add a chemical potential term, $\mu Q_B$.
Since $Q_B$ does not commute with generators $Q_{\pm}$, the chemical potential can drive the system periodically, which is not a contradiction.}
\begin{eqnarray}
\begin{pmatrix}
J_1^t, &J_1^x, &J_1^y
\end{pmatrix}
&=&
\frac{v^2\sqrt{2\pi\rho_B}}{\sqrt{\abs{k}}}
\begin{pmatrix}
0, &\cos({2\omega t}), &-\sgn(k)\sin(2\omega t)
\end{pmatrix}\\
\begin{pmatrix}
J_2^t, &J_2^x, &J_2^y
\end{pmatrix}
&=&
\frac{v^2\sqrt{2\pi\rho_B}}{\sqrt{\abs{k}}}
\begin{pmatrix}
0, &\sin({2\omega t}), &\sgn(k)\cos(2\omega t)
\end{pmatrix}.
\end{eqnarray}
Since this obviously breaks the rotational invariance, we can conclude that the ground state configuration breaks the rotational symmetry spontaneously.
In other words, the configuration of $SU(2)_k$ Chern-Simons theory coupled to fundamental scalar is homogeneous but anisotropic, at large baryon number.

If one further wants a Lorentz and global $SO(3)$ invariant, (sufficient) criterion for the anisotropy, that would be
\begin{equation}
\epsilon_{\mu\nu\sigma} J^\mu_{B} J^\nu_1 J^\sigma_2 \neq 0
\end{equation}
What is also interesting about the symmetry breaking pattern of this system at large charge is that although both the spatial rotation and the global $SO(3)$ symmetry are spontaneously broken, a combination of both is not.
The analysis of such a breaking pattern will not be further analysed here, but this is an important future work.

\section{Discussion of the result}
\label{consequence}

\subsection{Regime of validity at large $k$}
\label{sec:validity}

Since we have two tunable parameters, $k$ and $Q$, in the system, we might worry that the validity of the EFT \eqref{eq:EFT} if we take large $k$ in relation to $Q$.
We will argue that this is not the case here, however, for both critical boson and regular boson cases.
We will assume $k>0$ hereafter, without loss of generality.

\subsubsection{Critical boson theory}

For the critical boson theory, we expect that the value of $h$ is always of $O(1)$, even when the limit $k\to \infty$ is taken, since the theory is never connected to a free theory in the bosonic description.
The large-$k$ behaviour of the ground state configuration is the following,
\begin{align}
    \abs{\phi}^2&=O(\sqrt{\rho}), \quad \omega=O(\sqrt{\rho})\\
\begin{pmatrix}
A^t, & A^x, & A^y
\end{pmatrix}
&=
\begin{pmatrix}
-O(\sqrt{\rho}/k) \times T^{(3)},&
O(\sqrt{{{\rho}}/{{k}}})\times T^{(1)},&
O(\sqrt{{{\rho}}/{{k}}})\times T^{(2)}
\end{pmatrix}    
\end{align}

By expanding around the VEV, \eqref{eq:vev}, and looking at the part where it is quadratic in fluctuation, we can deduce the size of the fluctuation of the fields.
Let us first take the temporal gauge, $A^t=0$.
For the $\abs{\phi}$ mode, the term quadratic in terms of fluctuation is of order $k^0\rho_B$, so the fluctuation is suppressed by $k^0/\sqrt{\rho_B}$.
This means that our result for the ground state configuration of $\phi$ can be trusted up to order $O(1/\sqrt{\rho_B})$, irrespective of the value of $k$.

Likewise, from the coupling between the gauge field and $\phi$, the fluctuation term in terms of $A^{x, y}$ is order $k^0\sqrt{\rho}$, while from the Chern-Simons term the fluctuation term scales as $k$.
This means that the fluctuation of the gauge fields are suppressed by the smaller of  $k^0/\sqrt[4]{\rho}$ or $1/\sqrt{k}$.
Therefore, when $\rho\gg k^2$, the fluctuation goes as $1/\rho^{1/4}$, while when $k^2\gg\rho$, the fluctuation goes as $1/\sqrt{k} \ll 1/\rho^{1/4}$, and for both cases the fluctuation is suppressed at large $\rho$.
We therefore conclude that our computation of the ground state configuration can be trusted at large $\rho$, irrespective of the value of $k$.

\subsubsection{Regular boson theory}

For the regular boson theory, we expect that the value of $h$ is order $O(1/k)$ at large $k$ \cite{Aharony:2011jz}. 
Theory is connected to a free theory in the bosonic description at large $k$.
We denote $hk=\lambda$, which we expect to be $O(1)$.
The large-$k$ behaviour of the ground state configuration in this case is the following,
\begin{align}
    \abs{\phi}^2&=O(\sqrt{\rho k}), \quad \omega=O(\sqrt{\rho/k})\\
\begin{pmatrix}
A^t, & A^x, & A^y
\end{pmatrix}
&=
\begin{pmatrix}
-O(\sqrt{\rho/k}) \times T^{(3)},&
O(\sqrt{{{\rho}}/{{k}}})\times T^{(1)},&
O(\sqrt{{{\rho}}/{{k}}})\times T^{(2)}
\end{pmatrix}    
\end{align}

The following the same logic as in the critical case, we see that the fluctuation of $\phi$ is the smaller of $O(1)$ and $O(\sqrt{k/\rho})$, while that of the gauge fields is the smaller of $O(1/\sqrt{\rho k})$ and $O(1/\sqrt{k})$.
This means that, either way, the fluctuation of the gauge fields is suppressed by $O(1/\sqrt{\rho k})$, while that of the scalar is bounded above by $O(1)$.
We therefore conclude that the result of the ground state configuration is suppressed at large $\rho$.
Note that the strict $k\to\infty$ limit is however problematic, since $\abs{\phi}$ goes to infinity and thus experiences the runaway.
This is the same behaviour observed in \cite{Watanabe:2019pdh}, but this time without the conformal coupling since the theory is put on the torus.
Because of the absence of the conformal coupling, we cannot stabilise the negative mass term created by the chemical potential.

\subsection{Comparison with $O(4)$ Wilson-Fisher theory}

\subsubsection{Large-$k$ limit of Chern-Simons-matter theories}

The Chern-Simons-matter theories reduce to the ungauged models projected onto the singlet sector on the spherical spatial slice.
The argument goes as follows.
The Gauss law relates the flux of the gauge field and to the gauge charge as
\begin{align}
    \mathbb{Z}\ni \int_{S^2} F_{12}=\frac{1}{k}\int_{S^2} J^0.
\end{align}
This means by Dirac quantization the gauge charge with respect to any generator of the gauge group must be a multiple of $k$, and this decouples the non-singlet sector \cite{Banerjee:2012gh}.
Incidentally, this is a central proposal of the CFT dual of Vasiliev gravity \cite{1110.4386}.

On the torus spatial slice, the situation is totally different. 
It is in general impossible to project to the singlet sector of a theory in a local and unitary way there, which can be proven using modular invariance \cite{Banerjee:2012gh, Banerjee:2013mca}.
The physical picture for the case in question ($SU(N)$ Chern-Simons-matter theory with level $k$) is that we get more and more light modes as we take $k\to\infty$, for example the holonomies around the spatial torus, and they make it impossible to decouple the gauge field from the system.

\subsubsection{Comparing the ground state configuration at large baryon number}

We now check that the $SU(2)_k$ Chern-Simons theory coupled to fundamental scalar cannot reduce to the singlet sector of $O(4)$ Wilson-Fisher theory on the torus spatial slice, which is expected to what the original theory reduces to on the spherical spatial slice.
In order to do this, we will compare the symmetry breaking pattern of the ground state at large charge, and see that they do not match.
We summarise the symmetry breaking pattern of the two theories below.
\begin{itemize}
 \item $O(4)$ Wilson-Fisher theory: In \cite{Alvarez-Gaume:2016vff}, the authors proved the non-existence of homogeneous ground state configuration for the $O(2N)$ Wilson-Fisher fixed-point, if we excite the charges associated with more than one Cartan generators at the same time.
Afterwards, the ground state configuration was explicitly computed and found to be inhomogeneous at large charge for the $O(4)$ Wilson-Fisher theory  \cite{Hellerman:2017efx, Hellerman:2018sjf}.
In particular, we have no homogeneous ground state configuration for the $O(4)$ Wilson-Fisher fixed-point at large baryon number on the singlet sector.
 \item $SU(2)_k$ coupled to fundamental scalar: According to \ref{sec:validity}, our ground state configuration at large baryon number is homogeneous for any $k$, in particular $k\to\infty$.
\end{itemize}
This establishes that the $SU(2)_k$ Chern-Simons theory coupled to a fundamental scalar does {not} become the $O(4)$ Wilson-Fisher theory projected onto the singlet sector, on the torus spatial slice.



As pointed out in
\cite{Banerjee:2012gh, Banerjee:2013mca}, the specturm of Chern-Simons-matter theory at large-$k$ on the torus contains $k^{\alpha>0}$ almost degenerate states with gap $O(1/k)$.
The inhomogeneous ground state found in \cite{Hellerman:2017efx, Hellerman:2018sjf} is most likely hiding in such a large degeneracy.

Meanwhile, on the spherical spatial slice, our Chern-Simons-matter theory needs to reduce to the singlet sector of the $O(4)$ Wilson-Fisher theory.
This means that the symmetry breaking pattern of the ground state solution must match between the two sides.
The ground state configuration of the Chern-Simons-matter theory on the sphere includes vortices at the north and the south pole, which we can see by explicitly repeating the process of Sec. \ref{sec:ground}.
Such vortices might indicate the inhomogeneity of the ground state solution, but it might not survive the quantization of vortices.
The detailed study of the ground state on the sphere spatial slice is left for future studies.

\section{Conclusions and Outlook}
\label{conc}

We have computed the ground state configuration of $SU(2)_k$ Chern-Simons theory coupled to a fundamental scalar at large baryon number, on the spatial torus.
We found a homogeneous but anisotropic ground state configuration, valid as long as we take the baryon number large.
We also characterised such an anisotropy using the persistent global $SU(2)$ current present in the ground state of the system at large baryon number.

Chern-Simons-matter theories with chemical potentials has not been much considered yet, but this paper proved it important --
A large chemical potential for global charge densities renders the low-energy dynamics entirely semi-classical, even when the theory is strongly-coupled.
This will be quite useful in quantitatively verifying various particle-vortex dualities conjectured in three dimensions.
By directly accessing the strongly-coupling regime by semi-classical analysis presented in this paper, one should be able to compute physical quantities in both sides of the dualities, as an asymptotic expansion in terms of the global charge.

With this goal in mind, in future work we hope to study the following.

\begin{description}
\item[Generalisation to $SU(N)$ Chern-Simons-matter theories]\mbox{}\\
One might wonder if going to higher ranks wouldn't change anything, but that might be too hasty.
Specifically, as there are no counterparts of the $SU(2)$ global currents which were present in the $SU(2)$ Chern-Simons-matter theory,
one does not know how to characterise the anisotropy of the ground state gauge invariantly.

\item[Effective field theory of Chern-Simons-matter theories at large charge]\mbox{}\\
Based on the ground state configuration we have presented in the main body of the text and by integrating out the massive degrees of freedom, one can obtain the effective action of the theory at large baryon number.
This will be important in computing physical quantities, not just at leading order, but to higher orders in the inverse charge expansion.
The computation should be of great use in analysing Chern-Simons dual theories on both sides and comparing physical quantities.

\item[{Symmetry breaking pattern at large charge}]\mbox{}\\
By introducing a chemical potential to the system, some of the global symmetries as well as the Lorentz symmetry are explicitly broken.
It will be interesting to know exactly which of the symmetries are broken as well as to know which combinations are actually preserved.
Considering various discrete as well as one-form symmetries could be useful too.
As the breaking pattern must match in both sides of the dualities, this should give a non-trivial check (other than anomalies) of the particle-vortex like dualities.

\item[{Studying fermionic duals and verifying Chern-Simons-matter dualities}]\mbox{}\\
By repeating the same analysis for the fermionic duals of Chern-Simons-matter theories, one should be able to verify Chern-Simons-matter dualities quantitatively by writing down the effective actions for both sides as an asymptotic expansion in terms of global charge.

Understanding systems with fermions at large charge could be difficult if the fermions form a fermi sea (one does not know how to renormalise it at finite volume), but once dualities are understood at large charge, one could hope to also understand such systems, by following similar steps.
See also \cite{1904.07885} for another attempt to study Chern-Simons theories coupled to fermion at non-zero chemical potential.
\end{description}

\section*{Acknowledgements}
\begin{sloppypar}
The author is grateful to Ofer Aharony, Matthew Dodelson, Nozomu Kobayashi, Simeon Hellerman, Shunsuke Maeda, Kantaro Ohmori, Domenico Orlando, Susanne Reffert and Sho Sugiura for discussions.
I also especially thank Simeon for encouraging me to finish this paper throughout, as well as Nozomu and Shunsuke for discussions at an early stage of this work.
The author acknowledges the support by JSPS Research Fellowship for Young Scientists.
This work is supported in part by JSPS KAKENHI Grant Number JP16J01143.
\end{sloppypar}

\bibliographystyle{JHEP} 
\bibliography{references,cond-mat}

\end{document}